\documentclass{article}

\usepackage{arxiv}

\usepackage[utf8]{inputenc} 
\usepackage[T1]{fontenc}    
\usepackage{hyperref}       
\usepackage{url}            
\usepackage{booktabs}       
\usepackage{amsfonts}       
\usepackage{nicefrac}       
\usepackage{microtype}      
\usepackage{lipsum}
\usepackage{graphicx}
\usepackage{algorithm}
\usepackage{algorithmic}
\usepackage{csquotes}
\usepackage[table,xcdraw]{xcolor}

\title{Multi-Perspective Semantic Information Retrieval}

\author{
  Samarth Rawal \\
  Department of Computer Science \\
  Arizona State University \\
  \texttt{samrawal@asu.edu} \\
   \And
 Chitta Baral \\
  Department of Computer Science \\
  Arizona State University \\
  \texttt{chitta@asu.edu} \\
}

\begin{document}
\maketitle

\begin{abstract}
Information Retrieval (IR) is the task of obtaining pieces of data (such as documents  or snippets of text) that are relevant to a particular query or need from a large repository of information. While a combination of traditional keyword- and modern BERT-based approaches have been shown to be effective in recent work, there are often nuances in identifying what information is "relevant" to a particular query, which can be difficult to properly capture using these systems. This work introduces the concept of a Multi-Perspective IR system, a novel methodology that combines multiple deep learning and traditional IR models to better predict the relevance of a query-sentence pair, along with a standardized framework for tuning this system. This work is evaluated on the BioASQ Biomedical IR + QA challenges.
\
\end{abstract}

\keywords{Information Retrieval \and Semantic Information Availability }

\section{Introduction}

\subsection{Information Retrieval}
Information Retrieval (IR) is an active area of research with significant downstream practical and research-related applications, especially when paired with Question Answering (QA). The goal of an IR system is to take in a query and locate from a data repository pieces of information that are most relevant to the query. Such a system therefore requires the ability to parse and interpret a query (employing methods from the subfield of QA), as well as to evaluate and rank the relevance of data in the repository with respect to the query. This work will specifically focus on IR in the textual domain.

Information Retrieval systems have considerable research and practical value. IR systems are used in NLP tasks like question answering, document classification, and automatic document summarization. From a practical standpoint, systems that can accurately locate relevant data can have many domain-specific uses. For instance, there is considerable real-world value in being able to accurately and automatically locate relevant scientific literature from a repository of millions of papers that can answer a particular clinical query. An effective IR system can offset considerable manual human review of data and allow users to focus on information that is relevant. From casual web searching to specific tasks like clinical study results lookup, there is significant interest and value in the development of accurate and reliable IR systems.

However, discussing such a system begs the question, \emph{what does it mean for a sentence or document to be relevant to a query?} After all, a piece of information's "relevance" to a query is rarely a simple binary value. For instance, two sentences may both be relevant to a query, but one could be \emph{more relevant} than the other by containing more complete information. Part of the objective of IR is to determine what constitutes "relevance" and how to locate information that matches this. 

The meaning and nature of a "relevant" document given a query can depend on the domain and the task for which IR system is being used. For instance, in the biomedical domain, if a medical professional submits a query about the efficacy of a specific drug on a particular illness, they would likely be looking for the results of clinical trials, rather than general information about a particular drug (which could be interpreted as being relevant to the query, but would not be relevant with respect to the likely intent behind the search). 

In this work, we present a semantic Information Retrieval system for the biomedical domain which incorporates elements from several NLP tasks such as Sentence Relevance, Semantic Textual Similarity, and Semantic Information Availability. Additionally, we present the Alternating Optimization framework, a methodology for effectively tuning the weights for a system consisting of multiple composite components for both Document and Sentence Retrieval and Ranking. We demonstrate the effectiveness of this system and training approach in retrieving and ranking relevant documents and sentences from the MEDLINE/PubMed Baseline through the BioASQ Challenge.

\subsection{BioASQ Challenge}
The BioASQ Challenge is a semantic indexing and question answering competition for the biomedical domain \cite{bioasqoverview}. In this work, we will be focusing on Task B, Phase A of this challenge, specifically on the document retrieval and sentence retrieval components of this challenge. The BioASQ dataset consists of a series of queries. For each query, up to 10 gold "relevant" documents with respect to the query are provided in the form of their PubMed URLs, and up to 10 "relevant" snippets of text with respect to the query (most commonly single sentences), which have been obtained from the gold documents are provided. A "document" consists of the title and abstract of a paper on PubMed. The dataset from which these documents are obtained is the MEDLINE/PubMed Baseline repository, which consists of roughly 28 million articles (where an article consists of a paper's document and abstract).

For this particular challenge, teams receive a list of queries. For each query, the system must first obtain up to 10 most relevant documents, and from those documents, up to 10 most relevant snippets of text. Thus, the snippet retrieval task can be considered a "downstream" task of the document retreival task. The results are evaluated on a variety of metrics that include the F1 score and Mean Average Precision (MAP) of the document and sentences. 

Table \ref{relevant-nonrelevant-samples} provides examples of relevant and nonrelevant sentences, taken from documents in the MEDLINE/PubMed 2018 Baseline dataset, given a particular query. The samples in the table were obtained from the BioASQ Challenge, and "relevant" samples are defined as the Top 10 most relevant sentences from a particular query per the BioASQ Gold Dataset; the "nonrelevant" sentences are those outside the Top 10. 

\begin{table}[]
\begin{tabular}{|l|l|}
\hline
\textbf{Query} & What is the effect of TRH on myocardial contractility? \\ \hline
\textbf{\begin{tabular}[c]{@{}l@{}}Relevant\\ Samples\end{tabular}} & \begin{tabular}[c]{@{}l@{}}• Acute intravenous administration of TRH to rats with ischemic cardiomyopathy caused a \\ significant increase in heart rate \\• TRH can enhance cardiomyocyte contractility in vivo \\• TRH in the range of 0.1-10 mumol/l was found to exert a positive  inotropic effect on cardiac \\ contractility. \\• Thyrotropin-releasing hormone (TRH) improved mean arterial pressure (MAP) and myocardial \\contractility (dp/dtmax, -dp/dtmax, Vpm, and Vmax). \\• TRH improves cardiac contractility, cardiac output, and hemodynamics.\end{tabular} \\ \hline
\textbf{\begin{tabular}[c]{@{}l@{}}Nonrelevant\\ Samples\end{tabular}} & \begin{tabular}[c]{@{}l@{}}• These data suggest that 5-HT is an important transcriptional regulator  of the cardiac TRH gene. \\• The effects of thyrotropin-releasing hormone (TRH) and the TRH-analogs, 4-fluoro-Im-TRH \\(4-F-TRH) and, 2-trifluoromethyl-Im-TRH (2-TFM-TRH), on the cardiovascular system and \\prolactin (PRL) release were examined in conscious rats. \\• Thyrotropin-releasing hormone (TRH) has been shown to be scattered throughout the \\ gastrointestinal tract. \\• It is concluded that the enhancement by TRH of indomethacin-induced  gastric lesions is due to \\a combination of the central and peripheral actions of the ulcerogenic agents.\end{tabular} \\ \hline
\end{tabular}
\caption[Relevant and Nonrelevant Samples, Given a Query]{Relevant and nonrelevant samples, given a query.}
\label{relevant-nonrelevant-samples}
\end{table}

\section{Related Work}
\label{sec:headings}
\subsection{Traditional Information Retrieval Methodologies}
Variants of automated information retrieval systems have been implemented for the past several decades. At the heart of being able to retrieve documents is creating effective representations of them. Historically, models such as the Vector Space model and Probabilistic model, which rely on factors including term frequency, inverse document frequency, document length, have been used to come up with document representations \cite{mitra2000information}.

Once the documents have been converted from text into a numerical or vector-based representation, a \emph{ranking function} needs to sort the documents in terms of relevance to a particular query. The Okapi BM25 ranking function, a derivative of the Probabilistic retrieval model, is one of the most well-known and widely-used ranking functions in IR, and is a bag-of-words retrieval function that also relies on term frequency and inverse document frequency (TF-IDF).

\subsection{Neural Breakthroughs in NLP}

BERT (Bidirectional Encoder Representations from Transformers) is a language representation model that, along with similar types of models, has empirically been shown to yield state-of-the-art results in several NLP tasks, such as on the GLUE benchmark, and on tasks related to IR like Question Answering \cite{devlin2018bert}. BERT and related models are able to achieve significant performance gains in part due to the large corpora of data they are trained on. Moreover, the Transformer neural network architecture \cite{vaswani2017attention}, which BERT is built on, facilitates deep relationships between individual tokens in an input, thereby allowing the model to gain better contextual awareness of the relationships present in the input.

\subsection{Recent Semantic IR Systems}
Apache Lucene is an open-source search library which is widely used for developing and deploying IR systems, both in research and commercial applications \cite{lucene2010apache}. The software is able to turn raw data repositories into indices containing the data represented in a form which allows for efficient searching and ranking. Moreover, common ranking and searching algorithms, such as the aforementioned Okapi BM25 algorithm, are implemented in the software. Anserini is an IR toolkit built on top of Lucene that facilitates easier experimentation and reproduction of results when using bag-of-words ranking models like BM25 \cite{yang2018anserini}. Anserini also has a Python interface named Pyserini.

Recently, several IR systems have been developed that take advantage of the ability of BERT to capture contextual information. Because running BERT models is computationally expensive compared to ranking algorithms like BM25, these systems generally use BM25 or similar algorithms to perform a "coarse" level retrieval to narrow down candidates from the order of millions of documents to tens or hundreds, and then carry out finer reranking via BERT.

One such system is Birch, which utilizes a combination of Anserini for coarse ranking and BERT for fine ranking \cite{yilmaz2019birch}. BERT performs top Document Retrieval via Anserini and further re-ranking of the document list via BERT.

One key insight made in Nogueira et al.  \cite{nogueira2019passage}, which will be explored further in the context of the work detailed in this paper, is the notion that the Semantic Rank of a particular document can be represented by the weighted sum of the document's three most relevant Sentences. In this way, a basic formulation of Document Ranking via Sentence Ranking can be made. This idea will be discussed further below.

\subsection{Semantic IR in the Biomedical Domain}
There have been many innovations specifically in the biomedical and clinical NLP domains. Some widely-used pretrained and finetuned variants of BERT are BioBERT \cite{10.1093/bioinformatics/btz682biobert} and NCBI BlueBERT \cite{peng2019transferbluebert}. Various participants in BioASQ Challenges have utilized variants of BERT, such as BioBERT or BERT-Large, for the Sentence Ranking components.

Other participants in the BioASQ Challenge, have also utilized BERT-based systems for document and/or snippet retrieval purposes. Traditionally, such IR systems have consisted of two independent modules, a document ranking and sentence ranking module, with the overall system operating as a "pipeline". This formulation matches the BioASQ Challenge, where Sentence Retrieval is a downstream task of the Document Retrieval challenge. However, some participants, such as the team from Athens University of Economics and Business, have developed methodologies to jointly rank and retrieve documents and sentences, to successful results \cite{pappas2019aueb}. As utilizing BERT and BERT-like neural models can be computationally expensive, efficient ranking algorithms like the previously-mentioned BM25 algorithm are generally used to retrieve a smaller subset of potentially relevant documents from the full repository of about 28 million documents. The more contextual, computationally-expensive algorithms are then run on the smaller subset of documents rather than on the full repository.

\section{Multi-Perspective Biomedical Information Retrieval}
Having briefly covered the foundations of Information Retrieval systems, along with relevant prior work, we will discuss our IR system designed for Document and Sentence Retrieval in the biomedical domain. We have seen from previous literature that a combination of traditional and neural network-based approaches can incorporate greater contextual awareness in IR algorithms, leading to better results. In this section, we will describe our semantic Information Retrieval system, including the various iterations (both successful and unsuccessful) of the algorithm throughout the course of its development.

\subsection{Motivation Behind Multi-Perspective IR}

As detailed in previous work, training a BERT model on a Binary Query-Sentence Classification task, then using the prediction scores during inference time, can be an effective methodology to rank the relevance of sentences given a query. We found that employing methods similar to the existing work discussed in Chapter 2 provided decent results on the BioASQ Challenge. However, there are a few key issues that remain unaddressed with this method. The central problem with solely relying on a binary Sentence Relevance-based system for Sentence Ranking tasks is that \emph{binary Sentence Relevance is not equivalent to Sentence Ranking by importance}.
 
 There are several aspects of binary Sentence Relevance that make it advantageous for training such systems on. Primarily, there is a significant amount of labeled data for this task, both in general English and in biomedical domains. Moreover, it is relatively simple to generate additional data, both automatically and semi-automatically, by modifying existing datasets, as we did with the BioASQ dataset. However, intuitively speaking, just because a sentence is "relevant" does not mean it is a top-ranked sentence in terms of what a user submitting a query is looking for. In other words, "relevance" is generally not a binary attribute -- there are different degrees of relevance -- two sentences can both be relevant, but one can be more relevant than the other -- that cannot be adequately captured by such a binary formulation.
 
 Additionally, in the BioASQ challenge, sentences and documents are ranked by \emph{confidence of relevance} \cite{bioasqoverview}. One interpretation of this task is to rank sentences by the amount of "evidence" or "justification" that can be compiled about its relevance. This intuitive understanding leads us to the consideration of other metrics and methodologies that can be used to gather additional "justification" for the relevance of a sentence which can ideally fill the gaps present through the sole utilization of the Sentence Relevance-based methodology.
 
Thus, we propose the fusion of three models trained on different NLP tasks -- Sentence Relevance, Semantic Textual Similarity, and Semantic Information Availability -- to create a joint representation of the relevance of a sentence. We will first describe each of these tasks and the process used to train models for the respective tasks, then discuss the methodology of fusing these three representations into a single one. 

\subsection{Sentence Relevance}
As noted in previous literature, utilizing BERT-based architectures for Sentence Relevance can provide comparatively effective results for sentence relevance ranking. This work follows the approach detailed by Nogueira et al. in repurposing BERT as a Passage Reranker \cite{nogueira2019passage}. A BERT-Large model (with 24 layers, 16 attention heads, and 340 million parameters) is trained on a binary classification task of classifying a query-sentence pair as a "relevant" pair or "not relevant" pair. During inference, the prediction probability is taken as the Sentence Ranking score; sentences with higher scores are considered to be more relevant. As noted by the authors in \cite{nogueira2019passage}, this model was surprisingly effective across various domains; specifically, they demonstrated its effectiveness across general English training and Twitter domain IR performance \cite{nogueira2019passage}. 

We start with a BERT model that has been fine-tuned on a binary classification task of whether a particular sequence was relevant to a given query or not, using the Microsoft MS MARCO dataset \cite{nguyen2016msmarco}. The model weights for such a finetuned version of this model, detailed in the paper, have been made available by the authors on GitHub \cite{nogueira2019passage}. The authors have trained the model on a dataset of 12.8 million \texttt{(query, sentence)} pairs from the MS MARCO dataset.

Using the BioASQ training data, we create a biomedical binary sentence relevance dataset in the same format in order to further finetune the MS MARCO-finetuned model using data from the biomedical domain. We generate binary training samples using the training set and evaluate using binary samples generated from the development set. The training and development binary sentence relevance datasets are 221K and 55K samples, respectively. We further finetune the model on this domain-specific binary sentence relevance task, evaluating on the development set. Examples from this dataset are shown in Table \ref{tab:sentence-relevance-examples}.

We find that, as the model has already been finetuned once on the MS MARCO dataset, only a small amount of further finetuning -- under 1 epoch -- is necessary to achieve a peak in the model's performance on both the binary sentence relevance score and the BioASQ Sentence Ranking score. Each checkpoint of the model is evaluated on both these measures, and it is found that further finetuning for 1000 checkpoints on the biomedical sentence relevance dataset leads to a small gain in the model's performance compared to the MS MARCO "baseline" model. Moreover, it can be seen that there is a correlation between performance on the binary sentence relevance task -- evaluated using the F1 metric -- and the sentence ranking task -- evaluated using the Mean Average Precision, per the BioASQ Guidelines, in that a higher score on the binary sentence relevance task largely correlates with a higher MAP score in the BioASQ sentence ranking task (when the documents are fixed with the gold input). The hyperparameters used to finetune this model are described in Table \ref{tab:sentrel-hyperparameters}.

\begin{table}[]
\begin{tabular}{|l|}
\hline
\begin{tabular}[c]{@{}l@{}}( Is nintedanib effective for Idiopathic Pulmonary Fibrosis?,    In this review, we present the positive results of \\ recently published clinical  trials    regarding therapy for IPF, with emphasis on pirfenidone and nintedanib.,     1  )\end{tabular} \\ \hline
\begin{tabular}[c]{@{}l@{}}( Is nintedanib effective for Idiopathic Pulmonary Fibrosis?,   Results will be reported in the first half of 2014.,\\    0  )\end{tabular} \\ \hline
\begin{tabular}[c]{@{}l@{}}(Mutation of which gene is implicated in the familial isolated pituitary adenoma?,   Germline mutations in \\ the aryl-hydrocarbon interacting protein gene  are identified     in around 25\% of familial isolated pituitary \\ adenoma kindreds.,  1  )\end{tabular} \\ \hline
\begin{tabular}[c]{@{}l@{}}(Mutation of which gene is implicated in the familial isolated  pituitary adenoma?,   However, the exact \\ molecular mechanism by which  its disfunction promotes tumorigenesis of pituitary is unclear.,    0  )\end{tabular} \\ \hline
\end{tabular}
\caption[Examples from the Biomedical Science Relevance Dataset]{Examples from the biomedical Sentence Relevance dataset.}
\label{tab:sentence-relevance-examples}
\end{table}
\begin{table}[]
\centering
\begin{tabular}{|l|l|}
\hline
\textbf{learning rate} & 3e-06 \\ \hline
\textbf{seq\_len} & 128 \\ \hline
\textbf{weight\_decay} & 0.01 \\ \hline
\textbf{adam\_epsilon} & 1e-08 \\ \hline
\end{tabular}
\caption[Hyperparamters Used to Finetune Sentence Relevance BERT Model.]{Hyperparameters used to finetune Sentence Relevance BERT model.}
\label{tab:sentrel-hyperparameters}
\end{table}

\subsection{Semantic Textual Similarity}
Semantic Textual Similarity (STS) is a benchmark that evaluates how similar two sentences are two each other, often on a scale from 0.0 - 5.0, with 0.0 representing entirely dissimilar sentences, and 5.0 representing semantically identical sentences (although not necessarily lexically exactly the same).

One benefit STS systems can offer in the context of IR is the ability to resolve synonyms or semantic similarities present in medical jargon, clinical symptoms, or the like. For instance, two terms such as "migraine" and "headache", or "abdominal pain" and "stomachache", can be semantically very similar. However, because these terms are not lexically identical it can be possible for systems to miss these. Although techniques such as query expansion exist in traditional IR systems, they are not always effective in identifying \emph{all} related words and consequently can miss particular associations that may semantically exist. STS systems can help resolve these similarities and identify terms that could be potentially relevant to the query which otherwise may have been missed.

For the purpose of our system, as well as the following examples, we train a BERT Model (TODO: ADD SPECIFIC PARAMTERS) on the MedSTS Dataset (TODO: CITE/REFERENCE).

As a concrete example to demonstrate the effectiveness of STS systems in practical IR tasks, we take the following passage, excerpted from \cite{zimran2016long}:

\begin{displayquote}
"Long-term efficacy and safety results of taliglucerase alfa up to 36 months in adult treatment-naive patients with Gaucher disease. Taliglucerase alfa is an intravenous enzyme replacement therapy approved for treatment of type 1 Gaucher disease (GD), and is the first available plant cell-expressed recombinant therapeutic protein. Herein, we report long-term safety and efficacy results of taliglucerase alfa in treatment-naive adult patients with GD. Patients were randomized to receive taliglucerase alfa 30 or 60 U/kg every other week, and 23 patients completed 36 months of treatment... All treatment-related adverse events were mild to moderate in intensity and transient. The most common adverse events were nasopharyngitis, arthralgia, upper respiratory tract infection, headache, pain in extremity, and hypertension...."
\end{displayquote}

\textit{The abstract has been truncated for brevity.}

\begin{figure}[h]
\includegraphics[width=16cm]{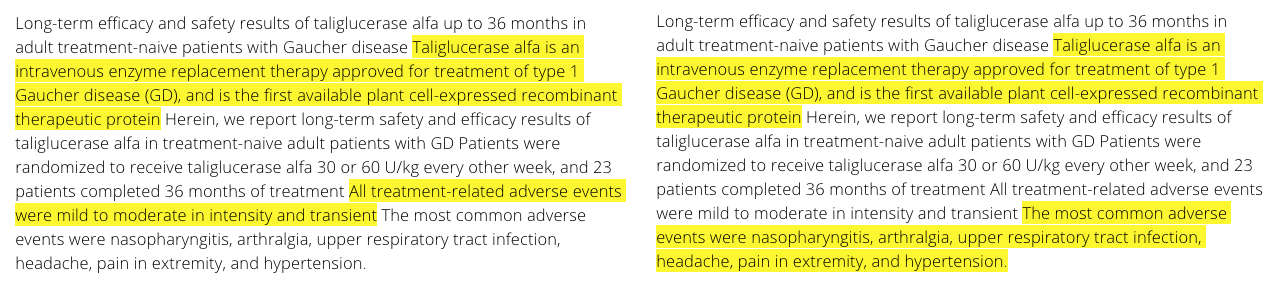}
\centering
\caption[Top 3 Ranked Sentences When Using Sentence Relevance Module (Left) and STS Module (Right).]{Top 3 Ranked Sentences when using Sentence Relevance module (left) and STS module (right).}
\label{fig:migraine-example-both}
\centering
\end{figure}

Given the above passage, and the following query:

\begin{displayquote}
"Can Gaucher disease treatment cause migraines?"
\end{displayquote}

a user would expect an IR system to return the last sentence in the excerpt, due to the fact that it mentions a connection between a potential treatment for Gaucher disease and headaches, and the semantic similarities between "migraines" and "headaches". However, when computing Sentence Ranking scores using the Binary Sentence Relevance model, we receive the highlighted sentences displayed in Figure \ref{fig:migraine-example-both} as the Top 3 ranked sentences. Thus, it is evident the Sentence Relevance-based module did not consider the potential semantic similarity between the two terms to be significant enough to warrant a high ranking score.

On the other hand, Figure \ref{fig:migraine-example-both} shows the Top 3 sentences when using the STS module to generate sentence rankings. In this case, the particular sentence recieved a high enough score to warrant a Top 3 ranking.

This example seeks to demonstrate the value of an STS system in a practical IR situation. As will be demonstrated in subsequent Sections and Chapters, utilizing solely an STS based system is not effective in a Sentence Ranking task; thus, there needs to be an integration of these multiple modules. 

For this task, a BERT-based system that had previously been finetuned on a clinical STS dataset was used.

\subsection{Semantic Information Availability}
\begin{table}[]
\centering
\emph{Given a query Q and a sentence S:}
\\
\begin{tabular}{|l|l|}
\hline
\textbf{Value} & \textbf{Description} \\ \hline
4 & Sentence S has the exact information related to query Q \\ \hline
3 & Sentence S has almost exact information related to query Q \\ \hline
2 & Sentence S has partial information related to query Q \\ \hline
1 & Sentence S has very little information related to query Q \\ \hline
0 & Sentence S has no information related to query Q \\ \hline
\end{tabular}
\caption[The Semantic Information Availability (SIA) Scale]{The Semantic Information Availability (SIA) scale}
\label{tab:sia-scale}
\end{table}

Existing scales used in NLP like Natural Language Inference (NLI) and Semantic Textual Similarity (STS) can be useful for finding and evaluting knowledge, like discussed previously. However, there is no standardized scale for information-seeking tasks that can answer the question \emph{Given a query Q and sentence S, how much information does S have that is needed to answer Q?} The motivation behind proposing a new scale, termed Semantic Information Availability (SIA), is to have a formal metric that can do so. Being able to evaluate query-sentence pairs in this manner can have direct practical benefits for tasks like IR, for which existing methodologies, as discussed previously, are not able to fully capture the necessities for effective ranking.   The SIA scale is defined in Table \ref{tab:sia-scale}, and further details about the scale are provided in \cite{rawal2020multi}.



As can be observed from the definition of the SIA scale, as well as the definitions of each of the components in the scale, this task is very aligned with the high-level goals of Information Retrieval -- to be able to score how well a particular sentence answers a query. Accordingly, the intuition is that having a module which, given a query and sentence as an input, predicts a score from 0 to 4 reflecting the amount of information that the sentence has to answer the query will be very useful for this Semantic IR task.

The major challenge in being able to utilize the SIA scale as part of a model is that there is no existing dataset. Consequently, it is necessary to create very large amounts of data (in the tens of thousands of entries) in order to train a system which can be an accurate predictor for SIA. In order to rapidly generate the data necessary to train a system on the SIA task, a largely automated methodology was taken that utilized existing datasets and generative models.

We use the methodolgy detailed in \cite{rawal2020multi} to generate a SIA dataset consisting of roughly 30,000 rows of $(query, sentence, sia\_score)$ tuples by adapting the QASC  Dataset \cite{khot2019qasc}:

The Question Answering via Sentence Composition (QASC) Dataset was used as a base for science-related SIA \cite{khot2019qasc}. This dataset consists of 9,980 8-way multiple-choice questions about grade school science, and comes with a corpus of 17M sentences. Each question is annotated with two facts from the corpus that can be combined together to arrive at the answer \texttt{(question, possible\_answers, correct\_answer, fact1, fact2, combinedFact)}. The following rules were used to convert this dataset into an SIA dataset:  

\textbf{For Category 4 Samples:} For each gold row in QASC, create SIA entry \texttt{(Q=question, S=combinedFact)}.

\textbf{For Category 2 Samples:} For each gold row, there are two potential SIA entries: \texttt{(Q=question, S=fact1)} and \texttt{(Q=question, S=fact2)}. However, because this is a multiple-choice task and SIA is meant to be an open-ended dataset, one of the answers will not be suitable for the SIA task as it will exploit the multiple-choice nature of the QASC dataset due to the data generation templates used for QASC. Thus, to select the best sentence, the BERT Sentence Relevance model finetuned on the MS MARCO dataset (described earlier this Chapter) was applied on both pairs, and the higher scoring pair was selected. This method was determined to be a useable heuristic by randomly hand-annotating around 50 samples and recognizing that this method correctly selected the useable sentence for each sample. 

\textbf{For Category 0 Samples:} Using a RoBERTa \cite{liu2019roberta} STS model trained on the GLUE STS-B general English task, two gold rows were picked in the QASC dataset, where the STS score between the two gold questions was above a particular threshold, determined heuristically. As long as the two queries did not match, two SIA rows were created: \texttt{(Q=question1, S=combinedFact2)} and \texttt{(Q=question2, S=combinedFact1)}.

Using this methodology, a preliminary SIA dataset in the science domain was created, consisting of roughly 30,000 rows of $(query, sentence, sia\_score)$ tuples.

From this initial dataset, two generative models are trained. Specifically, the OpenAI Generative Pretrained Transformer 2 (GPT-2) model is used \cite{radford2019languagegpt}. A GPT-2 Medium (355 Million parameters) is trained to generate a $query$ given a $sentence\_ 4$ value, and another is trained to generate a $sentence\_2$ given a $sentence\_4$. Both these models are trained only on the initially-generated dataset. These generative models enable creation of SIA samples in a purely unsupervised manner, rather than having to rely on existing datasets and repurposing them for the SIA task.

To generate Sentence 3 and Sentence 1 entries, the following method was used. For each Sentence 4, the ScispaCy library \cite{Neumann2019ScispaCyFA}, built on the SpaCy library \cite{spacy2}, was used to identify relevant biomedical entities from the sentence. Then, with a random probability, the entity is selected (and split into individual words if necessary). For each word, a substitution is found by finding the top 5 words by cosine similarity in a word2vec model trained on several clinical datasets, from which a random one is selected. The original word is swapped with its substitution. In this manner, a sentence from which part of the key entities are replaced with similar words is generated, and taken as Sentence 3 value. The same process is used to generate Sentence 1 entries; however, instead of using a Sentence 4 as the "base", the Sentence 2 is used.

Using both the generative models and word swapping techniques, a SIA dataset is constructed using sentences from the BioASQ training dataset. Roughly 200,000 SIA samples are created through this process. These samples are added to the first phase of auto-generated examples via the QASC dataset.

While the automatically-generated samples are comparatively less nuanced, and in some cases, less varied than examples which could be human-annotated, having an automated pipeline allows for the generation of the many examples needed to train a system from scratch.

Once the SIA dataset has been generated, a BERT-Large model is finetuned on this task, formulated as a regression task. The Mean Squared Error loss is used during the training of this model. During inference, a query and sentence are passed in to the model, which returns a score between 0.0 and 4.0, which is taken as the "SIA Score" for that particular query-sentence pair.

For application of the SIA module to the BioASQ Challenge, two versions of the model were developed; one using only the initial auto-generated data from QASC, and one using the data generated with the methodology described above. In practice, the former model was found to yield better results in the BioASQ Challenge.

While the current dataset and use for SIA is in a preliminary phase, the direct correlation between the components of the scale and real-world IR tasks make it a very promising ongoing area of focus to train better Semantic IR systems.

\section{Alternating Optimization}

One key objective for the Multi-Perspective Semantic IR system is how to weight the individual contributions of the subsystems to come up with the aggregate query-sentence score for ranking. To synthesize these "perspectives" we tried multiple approaches, including training feedforward layers to learn how to best synthesize outputs of the individual networks and utilizing a weighted sum-based technique. Further details about these experiments are found in \cite{rawal2020multi}. Ultimately, we found that utilizing a weighted sum-based approach to synthesize the scores generated from each of the perspectives into a single score yielded the best results experimentally.

Given that the overall goal for this system is to be applied on both Document and Sentence Ranking tasks, we propose a training framework that alternatingly optimizes between both these tasks.

Recall from Nogueira et al.\cite{nogueira2019passage} that a valuable proxy for the semantic relevance of a document is the weighted sum of its 3 most relevant sentences. Using this reasoning, we can consider the semantic relevance of a document to be a weighted sum of its BM25 retrieval score and the scores of its top 3 most relevant sentences:

\begin{center}
    $S_{DOC} =( \beta_{1} * S_{BM25}) +( \beta_{2} * \sum_{i=0}^{3}(w_i * S_{SENT\_i}))$
\end{center}

Similarly, we can consider the semantic relevance score of a particular sentence ($S_{SENT}$) to be a weighted sum of the scores predicted by the Sentence Relevance, Semantic Textual Similarity, and Semantic Information Availability modules, as well as the overall Document Score ($S_{DOC}$):

\begin{center}
    $S_{SENT} = \alpha_{1}S_{SentRelevance} + \alpha_{2}S_{STS} + \alpha_{3}S_{SIA} + \alpha_{4}S_{BM25}$
\end{center}

Note that Sentence Relevance, STS, and SIA scores are all scored on a different scale; their corresponding $\alpha$ values can be considered to be both a normalizing factor and a weighting factor.

\begin{figure}[h]
\includegraphics[width=9cm]{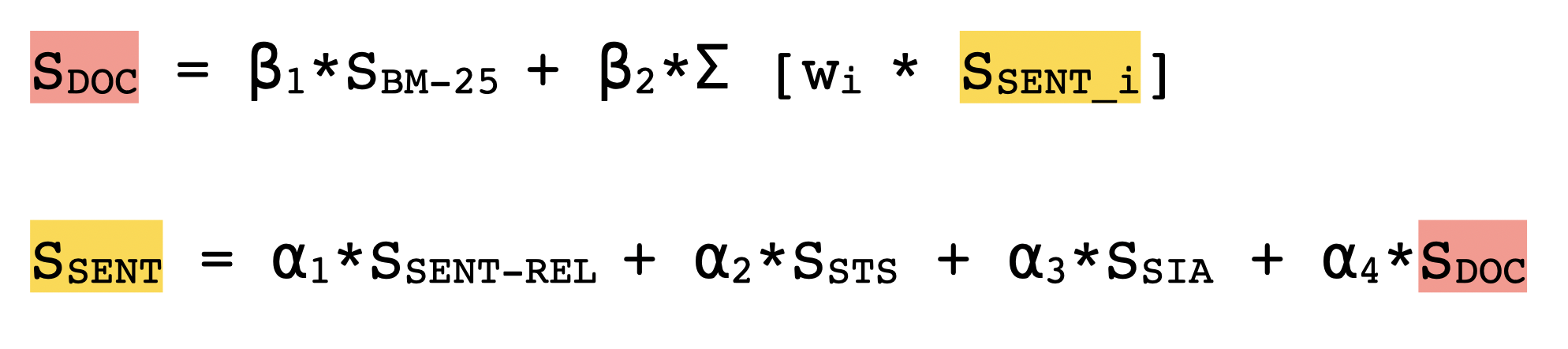}
\centering
\caption[Alternating Optimization Formulation of Document and Sentence Ranking.]{Alternating Optimization formulation of Document and Sentence Ranking.}
\label{fig:ao-formulation}
\centering
\end{figure}

In this manner, we have established a recursive relationship between the Document Score and individual Sentence Scores of the particular document, illustrated in Figure \ref{fig:ao-formulation}. 

In order to identify optimal parameters for these components, we utilize the following algorithm, first detailed intuitively and then algorithmically.

We first initialize all $ \alpha $ and $ \beta $ values equally. Next, we fix the Sentence Score parameters (the $ \alpha $ values) and use Bayesian Optimization to search for the optimal $ \beta $ parameters that maximize the Sentence Mean Average Precision score on the BioASQ training set. Once a predefined number of search steps have been carried out, the parameters for $ \beta $ with the corresponding highest Sentence MAP scores are retained and replace the previous $ \beta $ values. In the next phase, the $ \beta $ values are fixed, and the Sentence Score parameters ($\alpha$ values) are optimized over to also maximize the Sentence MAP. The corresponding $ \alpha $ parameters to the highest Sentence MAP obtained are then retained as the new $ \alpha $ values. This process is then repeated until convergence or for a predetermined number of loops.

To tune the Multi-Perspective System for the BioASQ Challenge, the training data provided for the BioASQ 6 challenge (consisting largely of data used in the BioASQ Years 1-5 competitions) was shuffled and split into training and validation data. The training data was used to finetune the BERT models, and the validation data was used in the Alternating Optimization process to maximize the Sentence MAP scores. The hyperparameters obtained through this process are listed in \ref{tab:hyperopt1-golddoc}.

\begin{table}[t]
\centering
\begin{tabular}{|c|c|}
\hline
\textbf{Hyperparameter} & \textbf{Value} \\ \hline
$\alpha_1$ Sentence Relevance & 0.6123 \\ \hline
$\alpha_2$ (Semantic Textual Similarity) & 0.2664 \\ \hline
$\alpha_3$ (Semantic Information Availability) & 0.0785 \\ \hline
$\alpha_4$ (Document Score) & 0.9879 \\ \hline
$\beta_1$ (BM-25 Score) & 0.0002 \\ \hline
$\beta_2$ (Sentence Relevance Score) & 0.8523 \\ \hline
$w_1$ (Top Sentence Score 1) & 0.9938 \\ \hline
$w_2$ (Top Sentence Score 2) & 0.0338 \\ \hline
$w_3$ (Top Sentence Score 3) & 0.0271 \\ \hline
\end{tabular}
\caption[Learned Hyperparameter Values Over BioASQ Sentence Relevance Development Dataset, with Gold Documents.]{Learned hyperparameter values over BioASQ Sentence Relevance development dataset, with gold Documents.}
\label{tab:hyperopt1-golddoc}
\end{table}

\section{Overall System Architecture}
\begin{figure}[h]
\includegraphics[width=11cm]{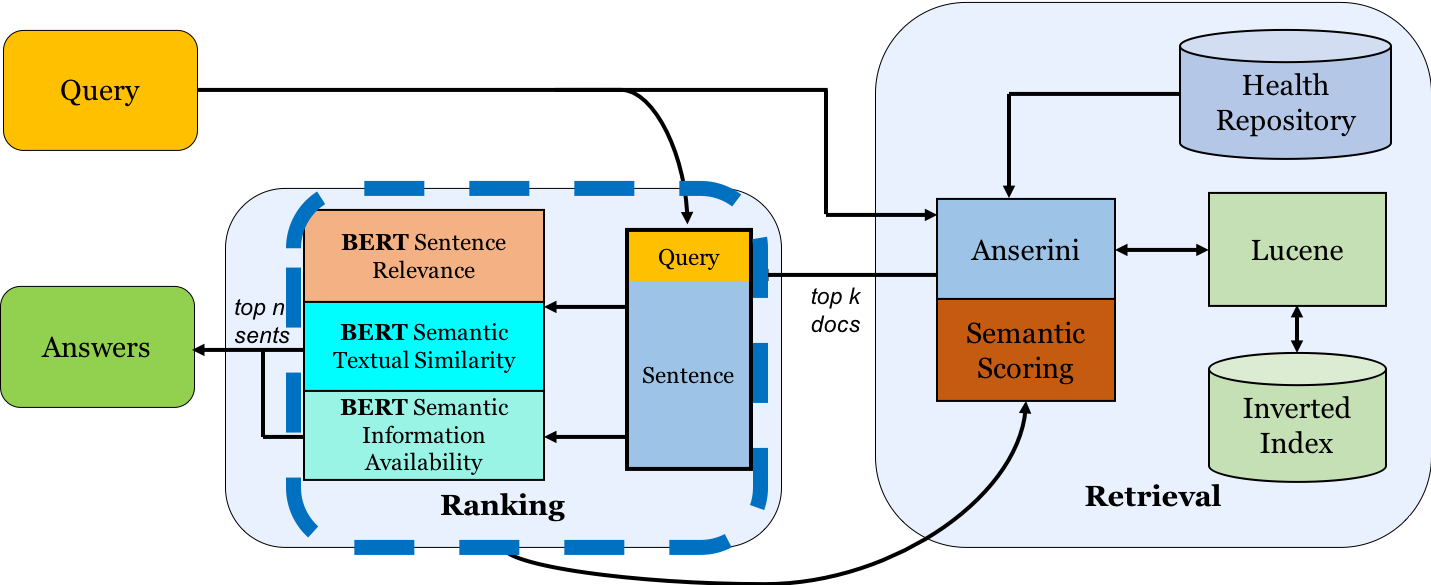}
\centering
\caption[Overall Semantic IR System Architecture]{Overall Semantic IR system architecture}
\label{fig:overallarchitecture-1}
\centering
\end{figure}

Figure \ref{fig:overallarchitecture-1} provides a high-level overview of the system as a whole as part of the BioASQ Challenge, with the Sentence Ranking module emphasized. The diagram illustrates the use of the Sentence Ranking module in both the Document Retrieval and Sentence Ranking tasks. The first task of the system for the BioASQ Challenge is to take a query and retrieve the top $k$ relevant documents out of roughly 28 million candidates in the MEDLINE/Pubmed Baseline, where $k$ is an integer that is set by the user depending on the objectives (for the BioASQ Challenge, $k$ is up to 10). Given the top $k$ relevant documents along with the original query, the second part of the system will break the documents down into individual sentences and perform semantic ranking on the sentences to identify the top $n$ most relevant sentences given the query, where $n$ is again an integer that is set by the user (for the BioASQ Challenge, $n$ is up to 10).

\section{Results} 
\subsection{BioASQ Document Retrieval Challenge}
The system described in this work was initially evaluated on the BioASQ 6B Phase A Document and Snippet ranking challenges, due to the comparatively larger amount of existing literature available regarding this challenge for comparison and analysis purposes. However, after evaluation on the BioASQ 6B results, this system was also evaluated on the BioASQ 7B Phase A Document and Snippet ranking challenges. Note that in the BioASQ Snippet Ranking component, a "snippet" is not necessarily identical to a "sentence" (it can be a portion of a sentence or multiple sentences); however, practically speaking, the majority of snippets were single sentences. Thus, for the purposes of this system, a snippet is considered equivalent to a single sentence.

\emph{Note:} For both the BioASQ 6B and BioASQ 7B Phase A competitions, there were 5 "batches" of test data which teams participated in. The tables shown in this section are the BioASQ 6B Phase A, Batch 2; and BioASQ 7B Phase A, Batch 4 results. These results are generally representative of those from the other batches; the full results for all batches are provided in \cite{rawal2020multi}.

The objective for the BioASQ Document Retrieval Challenge is, given a query, to locate the top $n$ documents from the MEDLINE/PubMed Baseline repository for the particular query, where $n$ is between 0 and 10 documents. The BioASQ 6 Challenge is based on the MEDLINE/PubMed 2018 Baseline repository, which consists of about 28 million documents, while the BioASQ 7 Challenge is based on the MEDLINE/PubMed 2019 Baseline repository, which consists of about 29 million documents. A "document" in the context of the BioASQ Challenge and MEDLINE/PubMed repository is the title and abstract of all completed citations in MEDLINE present in that year.

As mentioned earlier, the system was intended to be evaluated on the BioASQ 6B Challenge and was subsequently also evaluated on the BioASQ 7B Challenge. As a result, the system was trained using the BioASQ 6B Training Set, rather than the BioASQ 7B Training Set, which contains additional data. Consequently, re-training and hyperparameter tuning the system for the BioASQ 7B Challenge specifically may lead to improved results in that task, which is left as an exercise for future work.

The results for the BioASQ 6B Phase A Challenge, alongside some other teams' submissions to the challenges, are detailed in Figure \ref{tab:bioasq6-doc}. The results for the BioASQ 7B Phase A Challenge are detailed in Figure \ref{tab:bioasq7-doc}. The \underline{Semantic-Doc} submission is the system that has been trained using the Alternating Optimization algorithm with the objective to \emph{maximize Document Mean Average Precision}; the \underline{Semantic-Sent} submission is the system that has been trained using the AO algorithm with the objective to \emph{maximize Sentence Mean Average Precision}. When using documents for the downstream Sentence Ranking task, although the Document Ranking score is used in the formulation of the Sentence Ranking score, the rank of the document with respect to the other documents is explicitly not utilized; the more important factor is that the document made the Top 10 list, and as few as possible irrelevant documents made the Top 10 list. This can be reflected in the fact that for both BioASQ Challenges, the Semantic-Sent system had a higher Document F1 score (which does not account for ordering and only on presence or absence of the document), while the Semantic-Doc system had a higher Document MAP score.

\begin{table}[]
\centering
\begin{tabular}{|c|c|c|c|
>{\columncolor[HTML]{FFFC9E}}c |c|}
\hline
\textbf{System} & \textbf{\begin{tabular}[c]{@{}c@{}}MPrec\\ Docs\end{tabular}} & \textbf{\begin{tabular}[c]{@{}c@{}}MRec\\ Docs\end{tabular}} & \textbf{\begin{tabular}[c]{@{}c@{}}F-Measure\\ Docs\end{tabular}} & \textbf{\begin{tabular}[c]{@{}c@{}}MAP\\ Docs\end{tabular}} & \textbf{\begin{tabular}[c]{@{}c@{}}GMAP\\ Docs\end{tabular}} \\ \hline
ustb\_prir3 & 0.3121 & 0.6379 & 0.3396 & 0.2512 & 0.0639 \\ \hline
ustb\_prir4 & 0.3121 & 0.6379 & 0.3396 & 0.2512 & 0.0639 \\ \hline
aueb-nlp-4 & 0.3220 & 0.6431 & 0.3479 & 0.2500 & 0.0660 \\ \hline
aueb-nlp-2 & 0.3210 & 0.6420 & 0.3475 & 0.2470 & 0.0701 \\ \hline
testtext & 0.3201 & 0.6355 & 0.3464 & 0.2467 & 0.0634 \\ \hline
ustb\_prir1 & 0.3201 & 0.6355 & 0.3464 & 0.2467 & 0.0634 \\ \hline
ustb\_prir2 & 0.3221 & 0.6618 & 0.3519 & 0.2458 & 0.0795 \\ \hline
aueb-nlp-3 & 0.3160 & 0.6365 & 0.3423 & 0.2416 & 0.0646 \\ \hline
aueb-nlp-1 & 0.3060 & 0.6294 & 0.3332 & 0.2319 & 0.0560 \\ \hline
\underline{Semantic IR} & \underline{0.2684} & \underline{0.5697} & \underline{0.2952} & \underline{0.2085} & \underline{0.0324} \\ \hline
\end{tabular}
\caption[BioASQ 6B Phase A Document Ranking Results]{BioASQ 6B Phase A Document Ranking results.}
\label{tab:bioasq6-doc}
\end{table}

\begin{table}[]
\centering
\begin{tabular}{|c|c|c|c|
>{\columncolor[HTML]{FFFC9E}}c |c|}
\hline
\textbf{System} & \textbf{\begin{tabular}[c]{@{}c@{}}MPrec\\ Docs\end{tabular}} & \textbf{\begin{tabular}[c]{@{}c@{}}MRec\\ Docs\end{tabular}} & \textbf{\begin{tabular}[c]{@{}c@{}}F-Measure\\ Docs\end{tabular}} & \textbf{\begin{tabular}[c]{@{}c@{}}MAP\\ Docs\end{tabular}} & \textbf{\begin{tabular}[c]{@{}c@{}}GMAP\\ Docs\end{tabular}} \\ \hline
aueb-nlp-1 & 0.2541 & 0.6668 & 0.2998 & 0.2102 & 0.0316 \\ \hline
aueb-nlp-2 & 0.2531 & 0.6523 & 0.2992 & 0.2092 & 0.0279 \\ \hline
aueb-nlp-4 & 0.2481 & 0.6445 & 0.2948 & 0.2080 & 0.0268 \\ \hline
aueb-nlp-5 & 0.4537 & 0.6416 & 0.4580 & 0.1968 & 0.0291 \\ \hline
aueb-nlp-3 & 0.2401 & 0.6451 & 0.2857 & 0.1962 & 0.0282 \\ \hline
lh\_sys4 & 0.2230 & 0.6121 & 0.2695 & 0.1752 & 0.0186 \\ \hline
\underline{Semantic IR} & \underline{0.2170} & \underline{0.5867} & \underline{0.2592} & \underline{0.1777} & \underline{0.0176} \\ \hline
MindLab QA...  & 0.2080 & 0.5664 & 0.2463 & 0.1724 & 0.0121 \\ \hline
MindLab QA... & 0.2080 & 0.5664 & 0.2463 & 0.1724 & 0.0121 \\ \hline
MindLab QA...  & 0.2080 & 0.5664 & 0.2463 & 0.1724 & 0.0121 \\ \hline
\end{tabular}
\caption[BioASQ 7B Phase A Document Ranking Results]{BioASQ 7B Phase A Document Ranking results.}
\label{tab:bioasq7-doc}
\end{table}

\subsection{BioASQ Sentence Retrieval Challenge}
In the scope of the BioASQ Snippet Retrieval Challenge, the system must retrieve the top $n$ snippets from the top $k$ documents that have already been retrieved by the system, given a query. $n$ and $k$ are each integers between 0 and 10, and for the purposes of this system, a snippet is equivalent to a single sentence. Therefore, Snippet Retrieval is a downstream task of the Document Retrieval challenge; the more effective the Document Retrieval system is in identifying documents with relevant snippets, the more effective the Snippet Retrieval system will be.

\begin{table}[h]
\centering
\begin{tabular}{|c|c|c|c|
>{\columncolor[HTML]{FFFC9E}}c |c|}
\hline
\textbf{System} & \textbf{\begin{tabular}[c]{@{}c@{}}MPrec\\ Snippets\end{tabular}} & \textbf{\begin{tabular}[c]{@{}c@{}}MRec\\ Snippets\end{tabular}} & \textbf{\begin{tabular}[c]{@{}c@{}}F-Measure\\ Snippets\end{tabular}} & \textbf{\begin{tabular}[c]{@{}c@{}}MAP\\ Snippets\end{tabular}} & \textbf{\begin{tabular}[c]{@{}c@{}}GMAP\\ Snippets\end{tabular}} \\ \hline

\underline{Semantic IR} & \underline{0.3805} & \underline{0.3643} & \underline{0.3357} & \underline{0.3731} & \underline{0.0787} \\ \hline
aueb-nlp-5 & 0.3852 & 0.2976 & 0.2653 & 0.3187 & 0.0352 \\ \hline
MindLab QA...  & 0.2878 & 0.2307 & 0.1985 & 0.2736 & 0.0065 \\ \hline
MindLab QA...  & 0.2888 & 0.2298 & 0.1986 & 0.2695 & 0.0071 \\ \hline
MindLab QA... & 0.2888 & 0.2298 & 0.1986 & 0.2695 & 0.0071 \\ \hline
aueb-nlp-4 & 0.2873 & 0.2146 & 0.1850 & 0.2337 & 0.0231 \\ \hline
aueb-nlp-3 & 0.2746 & 0.2041 & 0.1749 & 0.2272 & 0.0210 \\ \hline
aueb-nlp-2 & 0.2768 & 0.2133 & 0.1826 & 0.2256 & 0.0236 \\ \hline
aueb-nlp-1 & 0.2716 & 0.2055 & 0.1749 & 0.2226 & 0.0202 \\ \hline
ustb\_prir4 & 0.2179 & 0.6188 & 0.2566 & 0.1731 & 0.0205 \\ \hline
\end{tabular}
\caption[BioASQ 6B Phase A Snippet Ranking Results.]{BioASQ 6B Phase A Snippet Ranking results.}
\label{tab:bioasq6-sent}
\end{table}

\begin{table}[h]
\centering
\begin{tabular}{|c|c|c|c|
>{\columncolor[HTML]{FFFC9E}}c |c|}
\hline
\textbf{System} & \textbf{\begin{tabular}[c]{@{}c@{}}MPrec\\ Snippets\end{tabular}} & \textbf{\begin{tabular}[c]{@{}c@{}}MRec\\ Snippets\end{tabular}} & \textbf{\begin{tabular}[c]{@{}c@{}}F-Measure\\ Snippets\end{tabular}} & \textbf{\begin{tabular}[c]{@{}c@{}}MAP\\ Snippets\end{tabular}} & \textbf{\begin{tabular}[c]{@{}c@{}}GMAP\\ Snippets\end{tabular}} \\ \hline
aueb-nlp-2 & 0.3254 & 0.4308 & 0.3048 & 0.3409 & 0.0344 \\ \hline
aueb-nlp-1 & 0.3209 & 0.4321 & 0.3018 & 0.3249 & 0.0281 \\ \hline
\underline{Semantic IR} & \underline{0.2959} & \underline{0.3414} & \underline{0.2568} & \underline{0.3030} & \underline{0.0151} \\ \hline
aueb-nlp-2 & 0.3254 & 0.4308 & 0.3048 & 0.3409 & 0.0344 \\ \hline
aueb-nlp-1 & 0.3209 & 0.4321 & 0.3018 & 0.3249 & 0.0281 \\ \hline
\underline{Semantic IR} & \underline{0.2959} & \underline{0.3414} & \underline{0.2568} & \underline{0.3030} & \underline{0.0151} \\ \hline
aueb-nlp-5 & 0.3256 & 0.4403 & 0.3010 & 0.2976 & 0.0379 \\ \hline
MindLab QA... & 0.2276 & 0.2857 & 0.2093 & 0.2214 & 0.0052 \\ \hline
aueb-nlp-3 & 0.2563 & 0.3581 & 0.2346 & 0.2213 & 0.0196 \\ \hline
aueb-nlp-4 & 0.2550 & 0.3325 & 0.2318 & 0.2173 & 0.0178 \\ \hline
MindLab Red... & 0.2168 & 0.2718 & 0.1982 & 0.2000 & 0.0067 \\ \hline
MindLab QA... & 0.2112 & 0.2317 & 0.1819 & 0.1931 & 0.0058 \\ \hline
MindLab QA... & 0.1998 & 0.2669 & 0.1865 & 0.1892 & 0.0064 \\ \hline
\end{tabular}
\caption[BioASQ 7B Phase A Snippet Ranking Results]{BioASQ 7B Phase A Snippet Ranking results.}
\label{tab:bioasq7-sent}
\end{table}

\section{Analysis}

In the Document Ranking task, the system generally placed below the Top 5 submissions in the BioASQ 6 and 7 Document Ranking challenges. However, in the Snippet Ranking task, the system generally placed around 1st and 4th on the BioASQ 6 and 7 Snippet Ranking challenges, respectively, in terms of systems and 1st and 2nd place, respectively, in terms of participants. Interestingly, although Snippet Retrieval is a downstream task of Document Retrieval, the Multi-Perspective IR system is able to outperform in the Snippet Retrieval task other systems which had higher scores in the Document Retrieval task. 

The Sentence Ranking module seems to perform very well with respect to the upstream Document F1 score compared to the other systems. Specifically, the Document F1 scores for the two systems outperforming this system in Snippet MAP in the BioASQ 7 challenge are 0.3409 and 0.3249, respectively. This system's Document F1 score in that challenge is 0.2592. This seems to imply that, given a higher performance in the Document Ranking task, the Snippet Ranking performance may come closer or exceed the other systems'. The comparatively high performance of the Sentence Ranking module in BioASQ 6, where the difference between this system's F1 Document Score and that of other top-performing systems was considerably smaller than in BioASQ 7, appears to support this idea. All in all, the value of introducing Multiple Perspectives when conducting Semantic Ranking, as well as incorporating the scores of the top sentences from a document that a sentence belongs to into that particular sentence's Ranking score, seems to be empirically validated through the Snippet Ranking performance in the BioASQ Challenge.

There is potential to improve the Document Ranking module, which as noted above will likely have the downstream impact of boosting the Sentence Ranking performance as well. Although the main focus in this work was developing a Sentence Ranking module, and formulating the Document Ranking task as partially a Sentence Ranking task, focusing on developing a Document Ranking module with a greater emphasis on document-specific features may lead to improved performance in this area. 

\section{Conclusion and Future Work}
The Sentence Retrieval Multi-Perspective approach was shown to provide empirically better results than using the simple Sentence Relevance-based BERT model approach. Future work may potentially involve incorporating additional "perspectives" into the model, as well as additional lexical and word-based features, into the overall Sentence Retrieval module.

The Document Ranking module was formulated partially as a Sentence Ranking task; consequently, no document-specific modules besides the BM25 scores were considered in the module. To this end, focusing on document-centric modules, perhaps analogous to the modules considered for the Sentence Ranking tasks, would provide greater support for the Document Ranking task and improve scores. To this extent, methods that fall under a neural network-based approach similar to the Feedforward Fusion method as well as a Weighted Sum-type approach can be considered, but on the scale of entire documents rather than individual sentences. Doing so may provide a more cohesive perspective on the relevance of documents and enable the system to make more informed ranking decisions on the document level.

The Semantic Information Availability module contributed little to the overall Semantic IR system; however, there is potential for improvement of the SIA systems by obtaining additional, higher-quality data. To this end, coming up with better heuristics and systems for the purpose of automatically generating SIA data will remain an area of future focus. With improved methodologies for automatically generating data, SIA systems will ideally continue improving in their predictions. Given the similarities between the SIA task and the overarching objective of an Information Retrieval system -- especially compared to that of existing NLP scales -- improving SIA performance may lead to gains in associated IR systems as well.

Using the Alternating Optimization framework was found to boost the scores of the Multi-Perspective system significantly on the randomly split development set used to evaluate our system during the training phase. Utilizing Bayesian Optimization across the parameter space in the Alternating Optimization training technique proved to be far more effective than utilizing grid search. However, the current implementation offers no guarantee the best parameters have been reached, either mathematically or empirically. Moreover, the importance of decisions such as initial parameter values, whether to perform early-stopping after a particular degree of convergence, and so on have yet to be considered. A thorough empirical evaluation of these different parameters will likely yield better insight into how this Alternating Optimization training loop can provide better sets of parameters for the Document Ranking and Sentence Ranking tasks. Additionally, simply spending more computation time in considering a greater number of hyperparameter combinations per training loop may yield immediate improved results.

\bibliographystyle{unsrt}  
\bibliography{references}  


\end{document}